\documentclass[]{spie} 
\usepackage[]{graphicx}
\usepackage{amsmath}
\usepackage{amsfonts}
\usepackage{amssymb}

\newcommand{\Eo}{E_{\rm open}}
\newcommand{\Ec}{E_{\rm closed}}

\newcommand{\no}{n_{\rm open}}
\newcommand{\nc}{n_{\rm closed}}
\newcommand{\po}{p_{\rm open}}
\newcommand{\pc}{p_{\rm closed}}
\newcommand{\ko}{k_{\rm o}}
\newcommand{\kc}{k_{\rm c}}
\newcommand{\kT}{k_{\rm B} T}
\newcommand{\Weff}{W_{\rm eff}}

\newcommand{\DEgate}{\Delta E_{\rm gate}}
\newcommand{\DEconf}{\Delta E_{\rm conf}}
\newcommand{\DEint}{\Delta E_{\rm int}}
\newcommand{\xmax}{x_{\rm max}}

\newcommand{\bea}{\begin{eqnarray}}
\newcommand{\eea}{\end{eqnarray}}
\newcommand{\be}{\begin{equation}}
\newcommand{\ee}{\end{equation}}

\title{Resonant effects in a voltage-activated channel gating} 

\author{Ewa Gudowska-Nowak\supit{a}, Bart{\l}omiej Dybiec\supit{a} and Henrik Flyvbjerg\supit{b}
\skiplinehalf
\supit{a}Marian~Smoluchowski Institute of Physics, Jagellonian University, Reymonta~4, 30--059~Krak\'ow, Poland;\\
\supit{b}Materials Research Department, Ris{\o} National Laboratory, Frederiksborgvej 399, DK--4000 Roskilde, Denmark
}

\authorinfo{Further author information:\\ E.G-N.: E-mail: gudowska@th.if.uj.edu.pl, Telephone: +48 (12) 663 5567, Fax: +48 (12) 633 40 79 \\
B.D.: E-mail: bartek@th.if.uj.edu.pl, Telephone: +48 (12) 663 5566, Fax: +48 (12) 633 40 79 \\
H.F.: E-mail: henrik.flyvbjerg@risoe.dk, Telephone: +45 4677 4104, Fax: +45 4677 4109 
}

\begin{document} 
\maketitle 

\begin{abstract}
The non-selective voltage activated cation channel from the human red cells,
which is activated at depolarizing potentials, has been shown to exhibit
counter-clockwise gating hysteresis. We have analyzed the phenomenon with the
simplest possible phenomenological models by assuming $2\times 2$ discrete
states, i.e. two normal open/closed states with two different states of ``gate
tension.'' Rates of transitions between the two branches of the hysteresis curve
have been modeled with single-barrier kinetics by introducing a real-valued
``reaction coordinate'' parameterizing the protein's conformational change.
When described in terms of the effective potential with cyclic variations of the
control parameter (an activating voltage), this model exhibits typical
``resonant effects'': synchronization, resonant activation and stochastic
resonance.
Occurrence of the phenomena is investigated by running the stochastic dynamics of the
model and analyzing statistical properties of gating trajectories.
\end{abstract}



\section{INTRODUCTION}
\label{sect:intro} 

Ion channels are proteins that allow selective conduction of ions across a
biological membrane. The transitions between an open (conducting) and a closed
(nonconducting) conformation of proteins forming the channel can be registered by
measuring movement of ions across the cell membrane.
The switching of ion channels between different states is brought about\cite{HOD,HI} by
energy derived from thermal fluctuations, the electrical field and (or) binding
the ligands. In particular, 
for voltage-gated ion channels discussed in this paper, kinetics between conducting
and non-conducting states is controlled by the trans-membrane potential.
The sequence of open and closed states can be measured in individual channels by
the patch clamp technique\cite{HI} to record the ionic current flowing through
the single channel. The resulting time series recording is a starting point for 
kinetic analysis and modeling. A common assumption for most of the models used 
in the study
of single channel gating is that the protein can dwell only in a few discrete
states and that the switching between these states is considered random with the
transition probabilities depending only on the present state, and not on the
previous history of the channel.\cite{CO} However, existing 
electrophysiological literature\cite{LAB,JACK,POUL,FULG,MERCIK,Varanda} points out existence of some degree of memory in the function of ion channels. The identification of
short- and long-term memory in single channel recordings could be then an
essential tool to discriminate between various models of channel kinetics and
interpretation of the data.

In relation to the channel memory a set of intriguing data have been reported by {\it Kaestner
et al.} who 
have studied\cite{POUL} voltage-activated cation channel in the human red 
blood cell (RBC) membrane,
demonstrating that the degree of activation at a given membrane potential
depended strongly on the prehistory of the channel. The gating exhibited
hysteretic-like behavior with the quasi steady-state deactivation and
activation curves displaced by about 25\,mV. The effect has been explained\cite{us} 
by evaluating a kinetic model 
for voltage-activated channel gating
that leads to a delayed response to a cycling external field in a way
similar to the change of magnetization in ferromagnetic materials 
under the influence of an external magnetic field. Alike memory-phenomena
have been also discovered in the ligand-gated NMDA receptor\cite{nowak}, 
in voltage-gated HERG-like channels\cite{penn} and in the gating
kinetics of the vanilloid receptor\cite{smith}
which is a thermo-sensitive non-selective cation channel 
activated by noxious heat. 
In order to explain hysteretic behavior observed in channel response
to external stimuli some of the authors\cite{penn} suggested the existence of an
ultra-slow inactivation process that effectively introduces a lag
in the feedback between voltage and gating.
Alternatively,\cite{smith} a periodic sensitization of the channel 
on repeated heat stimulation has been assumed.

Viewed as a nonlinear and time-delayed response of a system to the cyclical
variation of a control parameter, hysteresis is a fairly common phenomenon
observed in bistable dynamical systems.\cite{jung,mahato} 
In particular, it has been argued\cite{gage,mahato2,schulten} that the hysteresis
loop area can be a useful quantity for identifying critical values of field
parameters responsible for the onset of resonance-like response of the system and
therefore helpful in designing field parameters for an optimal control.
In the series of articles, Mahato, Shenoy and Jayannavar\cite{mahato,mahato2}
performed studies aimed to understand connection between the stochastic 
synchronization and stochastic resonance
(SR) in a bistable system exhibiting hysteretic response to the external
saw-tooth-type driving. They have shown that the hysteresis loop defined 
in such a system is a good measure of synchronization of passages from one well
to the other under the influence of a Gaussian white noise. Moreover, 
at a particular value of the noise intensity, the maximum of the
signal-to-noise ratio (SNR) is achieved thus demonstrating appearance of the
stochastic resonance, that in turn, can be correlated with the hysteresis 
loop area. These findings suggest that hysteretic voltage-dependent gating
may be closely related to documented observation of stochastic resonance in biological
channels.\cite{bezrukov}

Since the discovery of the SR phenomenon more than 20 years ago, the idea of stochastic
resonance has gained a lot of interest in diverse disciplines of science where 
its presence has been demonstrated repeatedly\cite{GAM} in versatile
applications. At the subcellular level, existence of the SR has been also
proven in the ion channels\cite{bezrukov} where the sensory transduction of the signal (ionic current) can be optimized by the presence of intrinsic ambient noise. 
The result has provoked a question whether the SR
is a collective effect occurring in some finite assemblies of channels or,
perhaps could be also realized in a single channel activity. Further
theoretical studies have revealed\cite{GO} that in a single Shaker potassium
channel the SR phenomenon can occur if the parameters that describe its
functionality are set within the regime where the channel is predominantly
dwelled in the closed state.

In this contribution we discuss occurrence of resonant effects in a model
double-well system that has been used to describe voltage-dependent gating
kinetics in the RBC cation channel.\cite{us} Section II presents basics of the
reaction pathway modeling for that case. The interplay of voltage driving and
thermal fluctuations are studied in an overdamped limit and lead to hysteresis,
stochastic resonance and resonant activation. All those phenomena can be
registered in the analysis of the channel gating kinetics, as discussed in
Section III. Section IV is devoted to a standard two-state approximation within
the Kramers scenario. Summary and closing remarks are presented in Section V.

\section{Bistability and reaction pathway}
\begin{figure}
\begin{center}
\begin{tabular}{c}
\includegraphics[height=7cm]{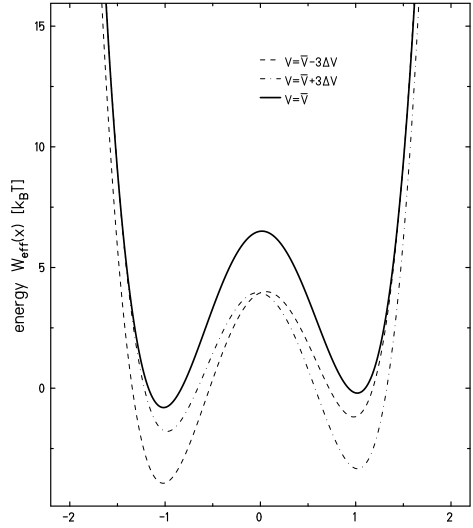}
\end{tabular}
\end{center}
\caption[example] 
{ \label{pot} 
The effective potential for conformational changes, $\Weff(x;V)$,
for $V=\bar{V}$, $V=\bar{V}+3\Delta V$ and $V=\bar{V}-3\Delta V$.
Parameters of the potential (in units of $\kT$) $\DEconf=0.6$, $\DEgate=9.72$,
$\DEint=2.16$, $\xmax=0.018$, $\Delta V=12$ and $\bar{V}=66$. The set of
parameters has been chosen \cite{us} to model hysteretic behavior observed 
in opening probabilities in the RBC channel.\cite{POUL}}
\end{figure}

The mechanism of voltage-gated ion channels was introduced in an early 
work of Hodgkin and Huxley who suggested that Boltzmann statistics
describes their probability of being open or closed, including the voltage-dependence
of this probability.\cite{HOD,HI}
For the purpose of modeling, we assume the ion channel in question has only
two states, open and closed.
We associate an internal energy $\Eo$, $\Ec$  with each of the two states, 
and assume that Boltzmann statistics describes
the probability for a channel to be open or closed,
with transitions between the two states 
being thermally activated barrier crossings
in an energy landscape that we do not detail.
Then the probability that a channel is open ($\po$) or closed ($\pc$) is
\be
p_{\rm open,(closed)} = \frac{ e^{-E_{\rm open,(closed)}} }{ e^{-\Eo} + e^{-\Ec} }
=\frac{1}{1+e^{\pm \Delta E}},
\label{eq:psigmagate}
\ee
where all energies are given in units of $\kT$.
In the case of voltage-gated channels, the transition 
between the open and the closed state possibly
involves movement of charges in the trans-membrane field, reorientation of
local polarization and conformational variation of channel protein that results in the
change of energy $\Delta E$ of the system.
With $\Delta E$ as above, Eq.~(\ref{eq:psigmagate}) 
may be written\cite{us}
\be
\po(V)
=\frac{1}{1+e^{-\alpha(V-\bar{V})}}\equiv f(V),
\label{eq:psigmaV}
\ee
where $\alpha \bar{V}$ replaces $(\Ec-\Eo)/\kT$, $\bar{V}$ is the voltage at which
$\po=\pc=1/2$ and $\alpha V$ parameterizes all other contributions to $\Delta E$. 
Equation~(\ref{eq:psigmaV}) may be derived by assuming Kramer's scenario 
for the gating kinetics.\cite{sig} This approach 
offers a description in which the configurational
change of the channel protein is modeled
by a multi-component ``reaction coordinate'' $x$. Since the potential energy of
the entire system (the gate, channel protein and membrane environment) is a function of
hundreds of
relevant coordinates of the system, $x$ can be understood as an effective
variable similar to its definition in the Marcus theory of charge-transfer
reactions.\cite{ulstrup}
The open and closed states of a channel can then be described
as $x\approx 1$ and $x\approx -1$, respectively, with a potential (activation)
barrier separating these two regions on the $x$-axis,
and $x$ being confined to the regions near $-1$ and $1$ by the potential taking
large values elsewhere. Opening and closing of a channel is then modeled
by a process similar to an activated uni-molecular chemical
reaction pictured by a kinetic scheme
\be
\mbox{open}
\begin{array}{c}
{}_{\kc(t)}\\
\rightleftharpoons\\
{}^{\ko(t)}
\end{array}
\mbox{closed.}
\label{di1}
\ee
In order to capture the configurational change of the protein quantitatively, we introduce
a parameterization for the reaction coordinate $x$ for this change by 
assuming that an equilibrium probability for the system to be in a conformational
state $x$ is given by
\be
p(x)=\sum_{\sigma=\pm1}p(\sigma,x)=e^{-\Weff(x;V)},
\ee
where $\sigma$ describes a binary variable $\sigma=\pm1$ discriminating between 
open ($\sigma=1$) and closed ($\sigma=-1$) states of the channel.
A corresponding effective potential (cf. Fig.~\ref{pot}) takes on the form\cite{us}
\be
\Weff(x;V)=W(x)-\log\left[ \cosh\left( \frac{1}{2}(\DEgate+\DEint x)\right) \right],
\label{weff}
\ee
where $W(x)$ describes the protein's internal energy as a function of $x$
\be
W(x) = \DEconf \left( \frac{1}{4} x (3-x^2) + \frac{3}{16\xmax}(1-x^2)^2 \right),
\ee
with $\DEconf$ being a measure of the difference in the internal energy
associated with the change of protein's configuration favoring open (closed)
states of the gate. The parameter $\xmax$ has been chosen as the value of the
reaction coordinate $x$ at which the local maximum, the top of the barrier
between $x=\pm 1$ states is located; $\DEgate=-\alpha(V-\bar{V})$
is contribution to the internal energy discriminating between the open (closed)
states of the gate and $\DEint=-\alpha \Delta V$ describes interaction energy between 
the gate and surrounding protein. In the above formula $\Delta V$ stands for the
maximum intensity of driving external voltage $V$ and $\bar{V}$ is an average
value of $V$ for which stationary probability $\po$ achieves value 1/2.

\begin{figure}
\begin{center}
\begin{tabular}{c}
\includegraphics[height=7cm,width=16cm]{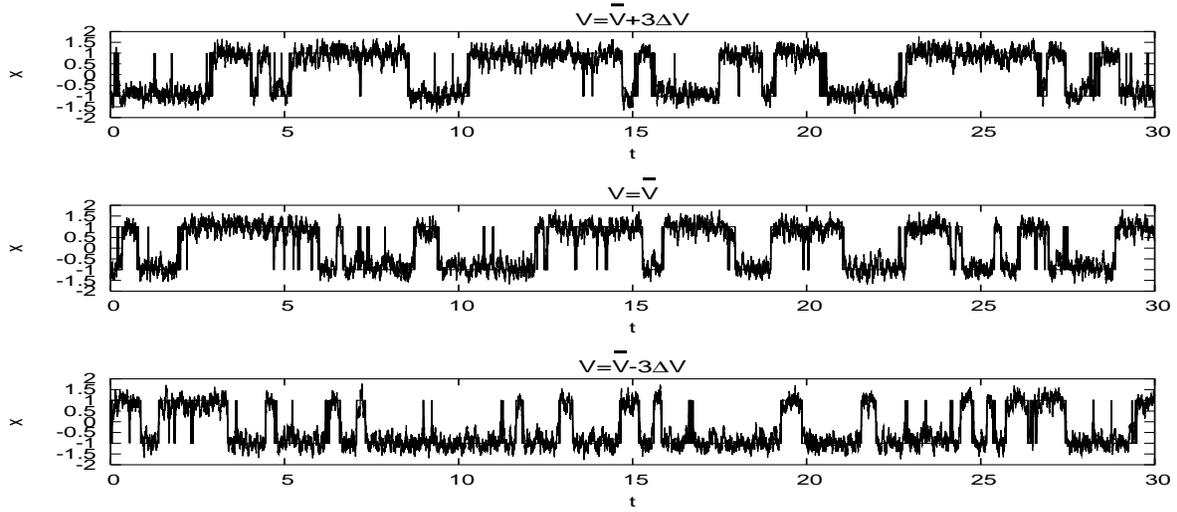}
\end{tabular}
\end{center}
\caption[example] 
{ \label{trajectory} 
Sample realizations of the stochastic process $X(t)$, for the two state and the continuous model,
derived from
numerical simulations of Eq.~(\ref{langevin}). Time step of simulations has been
set up to $10^{-3}$ with the frequency of the cycling voltage $\Omega=1$.
Trajectories have been run for the duration time $T\approx 10\pi$. Initial
value of $X$ has been sampled from the uniform distribution on the interval
$[-1.5,1.5]$.}
\end{figure}

We now additionally assume that the membrane potential
felt by a channel is not exactly constant, but changes,
albeit slowly so compared to the rates of the gate opening and closing
\be
V(t) = \bar{V} - \Delta V \cos\Omega t.
\ee
For $V(t)-\bar{V}$ large positive, stability analysis shows that 
conformational state $x=1$ is favored over $x=-1$ with an energy difference $\alpha\Delta
V=|\DEint|$. Similarly, for $V(t)-\bar{V}$ large negative
\be
\Weff(x;V)=W(x)+\frac{1}{2}\alpha(V-\bar{V}+\Delta V x)+\log 2
\ee
and the barrier at $\xmax$ becomes lowered by
$\frac{1}{2}\alpha|V-\bar{V}|=\frac{1}{2}|
\DEgate(V)|$ favoring state $x=-1$ over $x=1$. Note that for large differences 
$V(t)-\bar{V}$, addition of time-dependent driving does not change
characteristic relaxation frequencies of the system within the wells of the
potential $\Weff(x)$. Direct evaluation of
$\omega^2_{\xmax}=\Weff''(\xmax)$ and $\omega^2_{\pm 1}=\Weff''(x=\pm 1)$
yields $\omega^2_{\xmax}\approx -3\Delta E_c/4$, $\omega^2_{\pm 1}\approx 
3\Delta E_c/(2\xmax)$ for values of $\xmax$ close to 0. We further assume
that fluctuations of the collective variable $x$ quickly equilibrate, 
i.e. their dynamics can be well approximated by an overdamped Langevin equation
\be
\frac{dx}{dt}=-\frac{d\Weff(x;V)}{dx} + \xi(t),
\label{langevin}
\ee
with $\xi(t)$ representing effects of thermal noise modeled by uncorrelated 
Gaussian white noise, i.e $\langle\xi(t)\xi(t') \rangle=\sigma^2\delta(t-t')$. 
For convenience, we set the friction coefficient
$\gamma=1$ that rescales the time $t$ in the above equation.
Fig.~\ref{trajectory} presents samples of stochastic trajectories $X(t)$ simulated from
the above
Langevin equation at constant intensity $\sigma^2=8$ of the additive noise
(mimicking conditions of a constant temperature of the thermal medium). They model records of
current
in a biological channel switching between conducting and non-conducting conformations.
As expected from the former analysis, at large positive values of the gating voltage $V$, the
channel
is predominantly open with current fluctuating due to the local conformational variations
around the open
state ($x=1$). At $V=\bar{V}$ none of the states is favored, whereas for large negative $V$ a
preferential
stay in a closed state is visible. Increasing the noise intensity at a
constant value of the voltage cycling frequency causes a trajectory to fluctuate
wildly between both wells of the effective potential shortening the time
the process spends in the vicinity of the metastable points $x=\pm1$ and eventually,
destroying the pattern of the signal switching between ``open'' and ``closed''
states.

\begin{figure}
\begin{center}
\begin{tabular}{c}
\includegraphics[height=7cm,width=16cm]{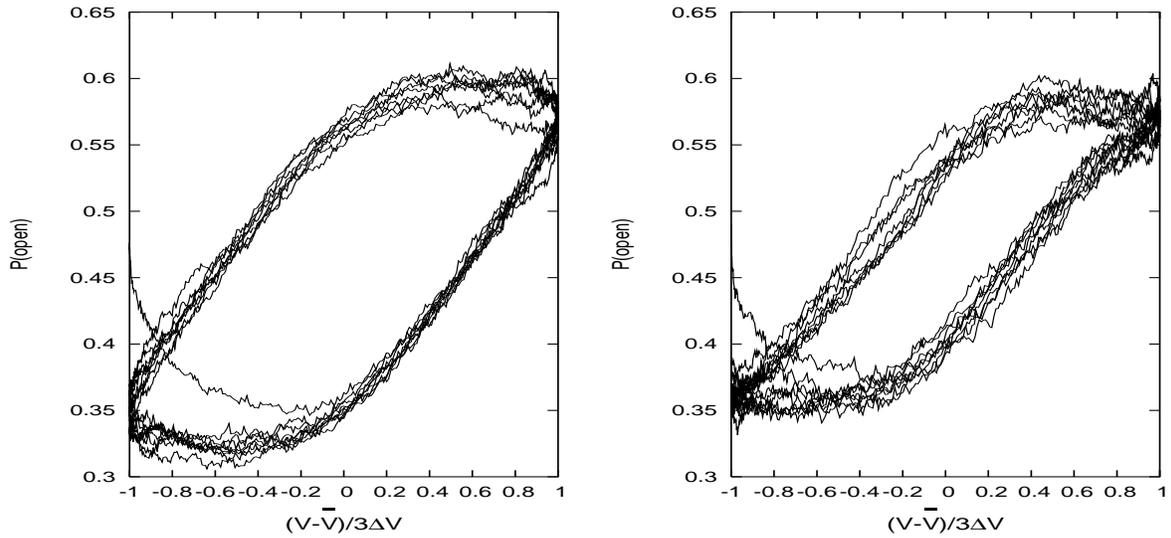}
\end{tabular}
\end{center}
\caption[example] 
{ \label{hist1} 
The open state probability $\po$ as a function of the membrane 
potential $V$. Probability has been estimated from the frequency histograms for 
$3\times 10^3$ trajectories mimicking passages between different conductance
levels. Intensity of the thermal noise $\sigma^2=6$ (left panel) and
$\sigma^2=8$ (right panel); time step of the simulation $dt=10^{-3}$.
}
\end{figure}

From the residence time frequency histograms for the open and closed states, we
have replotted probability of the channel being open as a function of the
driving voltage. Fig.~\ref{hist1} presents estimated opening 
probabilities for various intensities of thermal noise.

By increasing the
activating voltage from $V=\bar{V}$ to $V=\bar{V}+3\Delta V$ and reversing the
direction of voltage changes to $V=\bar{V}-3\Delta V$, an apparent hysteretic
shape of
the $\po(V)$ curves is observed (cf. Fig.~\ref{hist1}). The hysteresis changes its shape and area
with increasing intensity of the thermal noise and disappears for large values
of $\sigma^2$. 
Through cycling $V$, the height of the potential barrier dividing wells 
of open and closed states is now itself
a time-dependent variable. 
From other contexts\cite{GAM,ful,iwanisz,bezrukov}, it is known that 
this presence of periodic driving force and additional noise (fluctuations) may lead to
several
nontrivial resonant effects which we are now going to address in the forthcoming section.

\section{Dynamical hysteresis, stochastic resonance and resonant activation in channel gating kinetics} 

\begin{figure}
\begin{center}
\begin{tabular}{c}
\includegraphics[height=7cm,width=16cm]{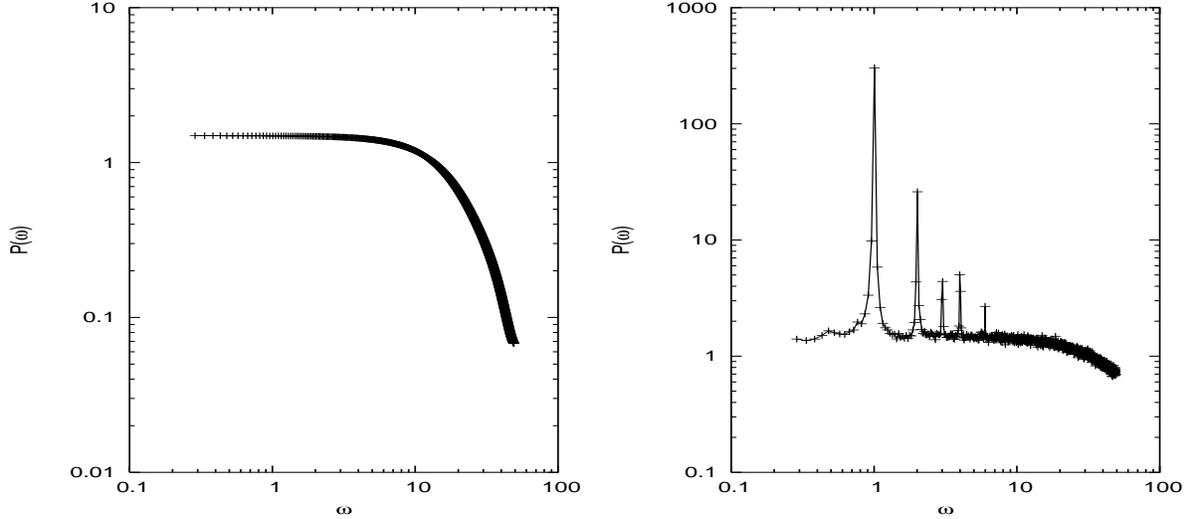}
\end{tabular}
\end{center}
\caption[example] 
{ \label{power1} 
Power spectra of the channel gating process 
for the two-state (left panel) and continuous model (right panel), as simulated from
Eq.~(\ref{langevin}) at the intensity of the additive thermal
noise $\sigma^2=0.2$ and $\Omega=1$. 
}
\end{figure}

\begin{figure}
\begin{center}
\begin{tabular}{c}
\includegraphics[height=7cm]{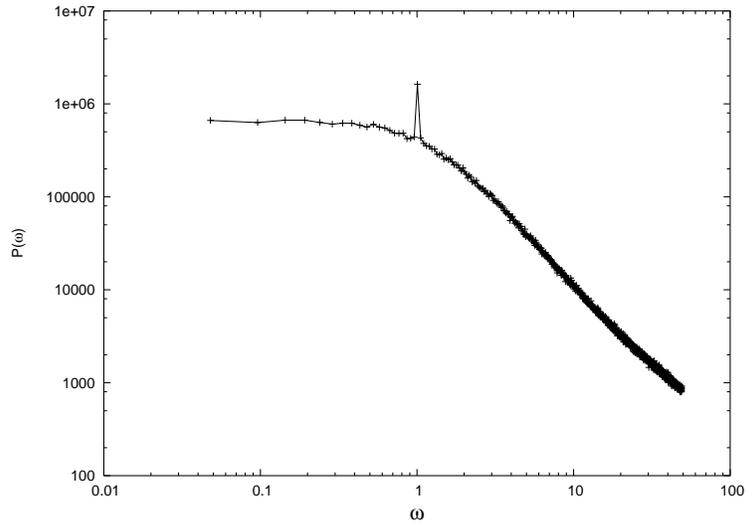}
\end{tabular}
\end{center}
\caption[example] 
{ \label{power2} 
Power spectrum of the channel gating process
for the continuous model simulated from
Eq.~(\ref{langevin}). Intensity of the additive thermal noise $\sigma^2=6$ and $\Omega=1$.
}
\end{figure}

An essential property of the stochastic process $\left\{X(t)\right\}$ is
interdependence of the random variable $X$ measured by the covariance function
$C(t,\tau)=\langle X(t+\tau)X(t)\rangle$ which through the Wiener-Khinchin theorem is
related to the power spectrum of the process $\left\{X(t)\right\}$. In the
framework of Fourier transformation, properties of the covariance function may
also be read off from its Fourier transform for a discretized version of
trajectories $X(t)$.
Figs.~\ref{power1} and~\ref{power2} present plots of the power spectral density as
a function
of frequency from a simulation of Eq.~(\ref{langevin}). The power spectra 
have been obtained from discrete periodograms for two different values of the
intensity of the additive noise $\xi(t)$. At low noise intensity 
$\sigma^2=0.2$, the power spectrum
exhibits sharp maxima at multiples of the unit frequency of the driving
voltage $V$ (cf. right panel of Fig.~\ref{power1}). 
The existence of spikes in the power spectrum is a general feature for
periodically driven systems \cite{jung3} and reflects (asymptotically) periodic 
character of the probability
distribution $p(x,t)$. This feature is absent in the power spectrum of
the two-state model (see left panel of Fig.~\ref{power1}). In this case 
any trajectory 
reaching the top of the potential
barrier at $\xmax$ is assumed to switch the system from one state to another,
so that the outgoing ``signal'' $X(t)$ is filtered and stored as a binary
series of $\pm 1$ values.
On increasing the noise intensity (or, equivalently temperature of the channel
environment) to $\sigma^2=6$, the structure of multiple maxima washes out
leaving the dominant peak at $\omega\approx\Omega=1$ observable for both - 
the two state approximation and the continuous model (cf. Fig.~\ref{power2}).
For large values of $\omega$, power
spectra scale to the $\omega^{-2}$ form reflecting exponentially decaying memory
of the gating process $\left\{X(t)\right\}$.

\begin{figure}
\begin{center}
\begin{tabular}{c}
\includegraphics[height=7cm]{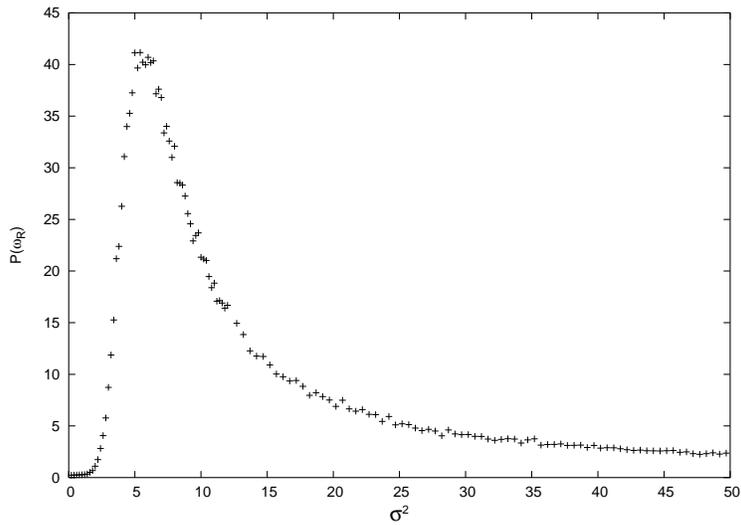}
\end{tabular}
\end{center}
\caption[example] 
{ \label{amp} 
Spectral amplification, in arbitrary units for the continuous model, of the signal (registered as a trajectory
passing from the left to the right well of the potential $\Weff(x;V)$) for
$\Omega=1$ at various noise intensities $\sigma^2$.
}
\end{figure}

The spectral amplification (an integrated power in the time-averaged power
spectral density at $\omega\approx\Omega=1$) has been calculated (cf. Fig.~\ref{amp}) according to the
interpretation due to Jung and H\"anggi.\cite{jung3}
The mean value of the $\left\{X(t)\right\}$ process has been obtained by
averaging over the ensemble of various realizations of the additive noise. For
long times $t\rightarrow\infty$ an asymptotic
trajectory looses memory of the initial conditions and becomes a periodic
function of time. Its square amplitude is then proportional to the spectral 
amplification expressed as the ratio of integrated power stored in the delta-like peaks
of the power spectra at $\pm\Omega$ to a total power of the modulation signal in
$V$.

\begin{figure}
\begin{center}
\begin{tabular}{c}
\includegraphics[height=7cm,width=16cm]{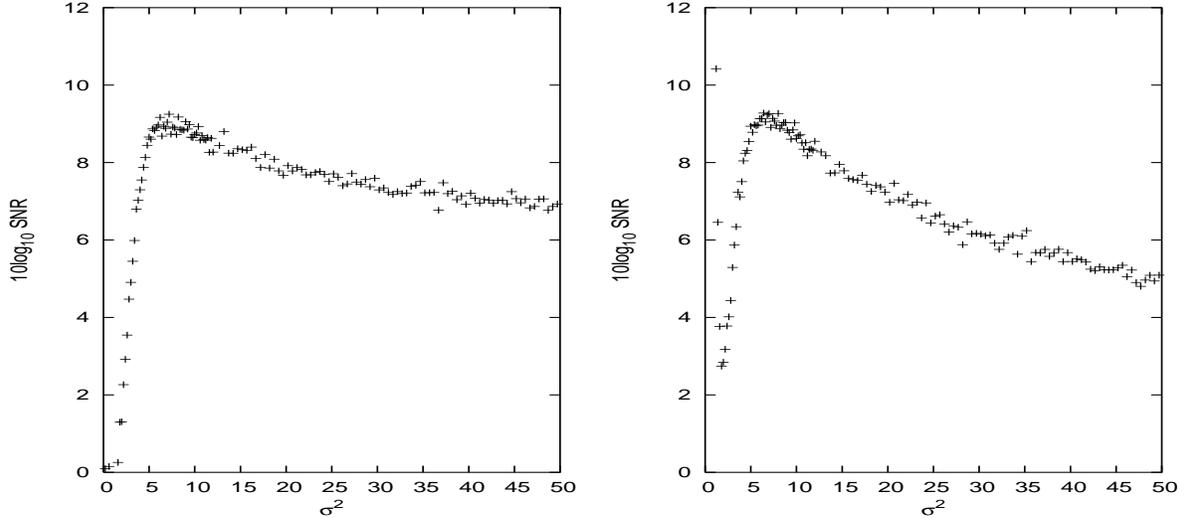}
\end{tabular}
\end{center}
\caption[example] 
{ \label{snr} 
The signal-to-noise ratio for $\Omega=1$ for the two state model (left panel) 
and for the continuous model (right panel), as a function of the noise strength 
$\sigma^2$.
The decaying part of the curve, in the right panel, represents relaxation processes in the potential well, details in the text.
}
\end{figure}

Spectral amplification displays a typical bell-shaped stochastic resonance
behavior as a function of increasing noise intensity with a
characteristic peak observed for
$\sigma^2\approx 6$ for both models under study. 
Similarly, evaluation of the signal-to-noise (SNR) ratio for the
potential under study produces curve (Fig.~\ref{snr}) that exhibits
non-monotonic resonant behavior for $\sigma^2\approx6$ and characteristic
divergence for the noise strength $\sigma^2\rightarrow 0$. 
For small $\sigma^2$ ($\sigma^2<2$) barrier crossing events are rare events,
and the decaying part of the SNR curve (cf. right panel of Fig.~\ref{snr})
corresponds to relaxation processes 
within the potential well. This feature is absent
for the two state model approximation.
In accordance with
the former studies of Mahato and Jayannavar\cite{mahato,mahato2}, the maximum
area of the hysteresis curve is observed for the noise intensity at which the
stochastic resonance is registered.

\begin{figure}
\begin{center}
\begin{tabular}{c}
\includegraphics[height=7cm]{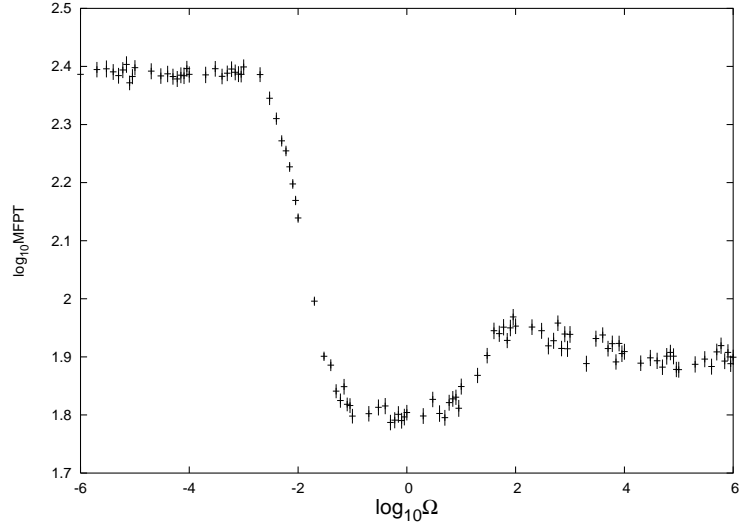}
\end{tabular}
\end{center}
\caption[example] 
{ \label{res} 
The MFPT for trajectories starting from the left minimum of the
potential $\Weff(x;V)$ as a function of driving voltage frequency $\Omega$ at 
the noise strength $\sigma^2=2$. Numerical results were obtained by use of Monte Carlo simulation of Eq.~(\ref{langevin}) with time step $dt=10^{-5}$ and averaged over $N=10^3$ realizations. Error bars represent deviation from the mean.
}
\end{figure}

The same system may also exhibit a different resonant behavior - so called resonant
activation (RA)\cite{doe,iwanisz} expressed by the shortest time of passage
over a fluctuating potential barrier in the presence of an additive noise.
As expected based on previous theoretical studies\cite{doe,iwanisz,bartek}, the RA occurs typically in
the bistable system under broad circumstances of varying shape of the potential
barrier and qualities of fluctuations. In the case of the modeled channel system, 
the resonant activation is brought up by an interplay of cyclic (deterministic)
variations of the barrier and an additive thermal noise (Fig.~\ref{res}).

\begin{figure}
\begin{center}
\begin{tabular}{c}
\includegraphics[height=7cm]{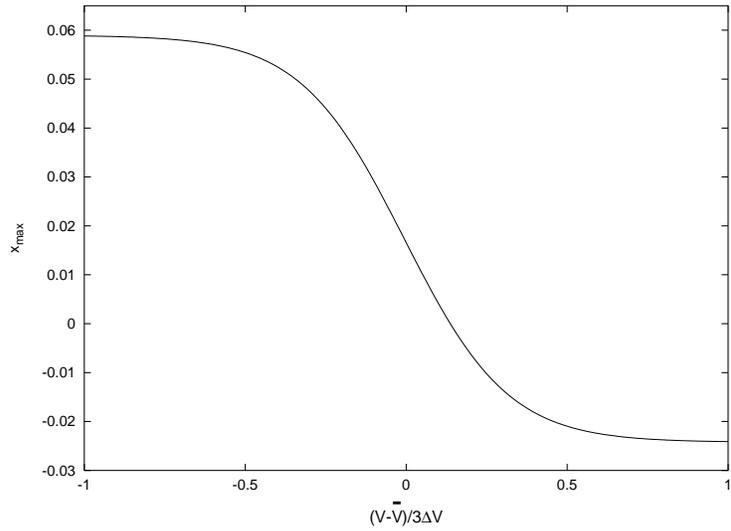}
\end{tabular}
\end{center}
\caption[example] 
{ \label{bor} 
Location of the barrier top for periodically perturbed potential as a function of the
driving voltage. Parameters $\DEconf$, $\DEint$, $\DEgate$, $\Delta V$ and $\bar{V}$ as in Fig.~\ref{pot}.
}
\end{figure}

In order to examine the statistics of the first passage times FPT for various shapes of
the potential $\Weff(x;V)$ introduced by the cycling voltage $V(t)$ own program
performing Monte Carlo simulations has been used. The trajectories Eq.~(\ref{langevin})
have been run starting from the left well until the top of the barrier has been hit,
where by the assumption, the trajectory has been absorbed. A position of the top of the barrier 
$\xmax$ depends on the driving voltage (see Fig.~\ref{bor}) and for a given set of the
parameters smoothly changes between $\xmax\approx 0.06$ and $\xmax\approx -0.03$.
Within the error limits of the simulations the resonant activation becomes visible at the noise
intensity $\sigma^2=2$ and the effect vanishes with increasing values of $\sigma^2$.

\section{Two state approximation}

In order to discuss the configurational change 
of the protein quantitatively 
we have introduced a parameterization of the reaction pathway
for this change. 
We now explore the consequences of the specific form of $\Weff(x;V)$, 
approximating the thermal barrier crossing with the simple Kramers 
scenario. In doing that, our description captures,
albeit approximatively only,
the consequences of a much larger set of potentials which are generic for
studying bistability and SR phenomena. 

In the overdamped Kramer's scenario\cite{Haenggi}, 
after assuming that the thermal energy $\kT$ is much smaller than the
respective barrier height $\Delta W^{\dagger}=\Weff(\xmax;V)-\Weff(\pm1;V)$, the
reaction coordinate undergoes a creeping motion and an adiabatic elimination of
the velocity 
$\dot{x}$ leads to the Langevin equation~(\ref{langevin}).
The time evolution of the corresponding probability $p(x,t)$ is governed by the
Smoluchowski equation. The spatial-diffusion rate in the potential $\Weff(x;V)$ can
be then evaluated by a steepest-descent approximation from the quadrature formulas
for the mean first passage time (MFPT)
\bea
\ko & = & k_{-1\rightarrow+1}(V(t)) \nonumber\\
& = & \nu_- e^{-(\Weff(\xmax;V(t))-\Weff(-1;V(t)))},\nonumber\\ 
\kc& = & k_{+1\rightarrow-1}(V(t))\nonumber\\
& = & \nu_+ e^{-(\Weff(\xmax;V(t))-\Weff(+1,V(t)))},
\eea
where the subdominant factor $\nu_{\pm}=(\omega_{\pm 1}\omega_{\xmax})/2\pi$ contains
product of the characteristic angular
frequency inside the metastable minimum ($x=\pm 1$) and an angular frequency at the top of
the potential barrier.

For a sufficiently high barrier separating open and closed states,
the probability of $x$ taking values in between vanishes and 
we can work with the integrated probability of $x$ being near $-1$ or $1$, 
$\nc(t)=\int_{-\infty}^{\xmax-\delta}p(x,t)dx$ and
$\no(t)=\int_{\xmax+\delta}^{\infty}p(x,t)dt$ where $\left\{-\delta, \delta \right\}$ 
describes an interval in the vicinity of the metastable $\xmax$.
The integrated probabilities satisfy $\nc(t)+\no(t)=1$ and the linear kinetic equation
\be
\frac{d\no(t)}{dt}=\ko \nc(t)-\kc\no(t).
\label{open}
\ee
Under stationary conditions the ratio of the reaction rates
can be related to the
difference in the energy by the equilibrium constant
\be
K=\frac{\kc}{\ko}=\frac{\nc}{\no}=\frac{\nu_+}{\nu_-}
e^{-[\Weff(-1;V(t))-\Weff(1;V(t))]}.
\label{constant}
\ee
Rearrangement of this equation gives a formula for the probability of
the channel being open Eq.~(\ref{eq:psigmaV})
\be
\po=\frac{1}{1+\kc/\ko}.
\ee
Thus $\Delta \Weff$ is related to the ratio of kinetic
rates for the ``forward'' and ``backward'' reactions. 
Experimental estimates of the
opening and closing rates can be found from the voltage-dependent dwell times
determined from experimental recordings
as inverse of averages
i.e. $\ko=\langle T_{\rm c}\rangle ^{-1}$, $\kc=\langle T_{\rm o}\rangle ^{-1}$
where $\langle T_{\rm c,o}\rangle $ are average dwell times in the closed (open) states.
The gating dynamics described by Eq.~(\ref{open}) can be otherwise considered as a
two-state Markovian process\cite{ful2,peter,lutz} with a time-dependent transition
probabilities between the states. A two state model of that type is known to describe
a stochastic resonance\cite{GAM,nam} and with both periodically and randomly modulated
rates has been investigated.\cite{lutz}.

The general solution of Eq.~(\ref{open}) can be readily found
\bea
\no(t)& =& e^{-[K_{\rm c}(t,0)+K_{\rm o}(t,0)]}\no(0) \nonumber\\
& + & C\int_0^t ds \ko(s) e^{-[K_{\rm c}(t,s)+K_{\rm o}(t,s)]},
\eea
where 
\be
K_{\rm o,c}(t,s)\equiv \int^t_{\rm s} dt' k_{\rm o,c}(t')
\ee
and 
\be
k_{\rm o,c}=\kappa_{\rm o,c}e^{-\alpha_{\rm o,c} V}[1+\alpha_{\rm o,c}\cos t +\lambda_{\rm o,c}\eta(t)],
\ee
where $\eta(t)$ stands for other sources of (nonthermal) noises of intensity 
$\lambda_{\rm o,c}$ that can perturb the transition rates. Note, that the Kramers rate
formula is derived under the assumption that the probability density within the well is
close to its equilibrium distribution, so that in order to use the modified Kramers rate
with a periodic perturbation of the voltage, its frequency must be much slower than the
characteristic rate for the probability to equilibrate within the well. Such an
adiabatic approximation is valid if the second derivative of the effective potential
$\Weff(x;V)$ at $x=\pm 1$, i.e. $\omega_{\pm 1}$ is much bigger than 1 (in our analysis
the characteristic period of the cycling voltage has been set up to 1). Parametric plots
of the cumulative open probability, as defined by $\no$ in function of $V$ display then the
dynamic hysteresis whose shape and loop-area depend on the ratio of the driving voltage 
frequency to the overall relaxation rate $k=\ko+\kc$. The hysteresis
disappears only in case of driving that has time scale significantly
different from 
the inverse of rates $k_{\rm o,c}$, i.e. when it is either very slow or very
fast. In such situations, the response of the system measured in terms of the
integrated probabilities $n_{\rm open,closed}$ depends on the applied voltage but
not on its frequencies (slow driving), or follows (fast driving) average, time
independent rates.
In consequence,
hysteretic behavior may be considered a generic for systems exhibiting the 
SR phenomenon. 
For the periodically driven two state process, recent analysis performed by Talkner\cite{peter2} 
has shown that studies of the entrance time densities in the system can 
detect the synchronization of the driving force and the driving process. 
However, a more detailed 
comparison of stochastic resonance and synchronization \cite{mahato2,peter2} shows that the effects are 
closely related but not identical.

\section{Summary}
There is no unique consensus in the SR literature on how to measure the SR effect. The
most popular performance measures used in the study depend on the forcing signal and
noise and can vary from system to system. The common approach to characterize SR is the
SNR that measures how much the system output contains the input signal frequency $\Omega$.
An adiabatic approximation\cite{GAM,nam} can then give an explicit SNR ratio for the
quadratic bistable potential. Alternative ways to measure SR are related to the
cross-correlation measures and estimates of the density probability of residence time\cite{mahato2} 
and escape rates.\cite{peter2}

In this work we have concentrated on a class of resonant effects that can be 
exhibited in the gating kinetics of the voltage-driven biological channel.
Our analysis reproduces all typical phenomena that are characteristic for the 
two-state system driven by fluctuations and periodic forcing. Albeit the latter
does not function here as an additive periodic force but
rather modulates the height of the free energy barrier that
discriminates between different conformational states of the gate, its
presence and interference with inherent thermal fluctuations leads to a 
fine-tuning of the gating kinetics exhibited by stochastic resonance and
resonant activation.

\acknowledgments 
The Authors acknowledge the financial support from the Polish State Committee for Scientific Research (KBN) through grants 1P03B06626 and 2P03B08225.


\end{document}